\documentclass[aps,nofootinbib,prd,eqsecnum,showpacs,showkeys,preprintnumbers,altaffilletter]{revtex4-1}

\usepackage{amsmath}
\usepackage{amssymb}

\makeatletter

\usepackage{graphicx}
\usepackage{amsfonts}
\usepackage{color}
\usepackage{bm}
\usepackage{mathrsfs}
\usepackage{epstopdf}
\usepackage{url}
\usepackage{footnote}
\usepackage{appendix}

\usepackage{float}
\usepackage{enumerate}
\usepackage[dvipsnames]{xcolor}

\usepackage{hyperref}

\usepackage{booktabs}
\AtBeginDocument{
\heavyrulewidth=.08em
\lightrulewidth=.05em
\cmidrulewidth=.03em
\belowrulesep=.65ex
\belowbottomsep=0pt
\aboverulesep=.4ex
\abovetopsep=0pt
\cmidrulesep=\doublerulesep
\cmidrulekern=.5em
\defaultaddspace=.5em
}

\usepackage{tabularx}

\newcolumntype{L}[1]{>{\hsize=#1\hsize\raggedright\arraybackslash}X}%
\newcolumntype{R}[1]{>{\hsize=#1\hsize\raggedleft\arraybackslash}X}%
\newcolumntype{C}[1]{>{\hsize=#1\hsize\centering\arraybackslash}X}%

\newcommand{\be}{\begin{equation}}
 \newcommand{\ee}{\end{equation}}
\newcommand{\ben}{\begin{eqnarray*}}
 \newcommand{\een}{\end{eqnarray*}}
\newcommand{\bea}{\begin{eqnarray}}
 \newcommand{\eea}{\end{eqnarray}}
\newcommand{\bdm}{\begin{displaymath}}
 \newcommand{\edm}{\end{displaymath}}
\newcommand{\ba}{\begin{align}}
 \newcommand{\ea}{\end{align}}

\makeatother

\begin{document}

\title{Quantum cosmology of Eddington-Born-Infeld gravity fed by a scalar field: the big rip case}

\author{Imanol Albarran $^{1,2}$}
\email{imanol@ubi.pt}

\author{Mariam Bouhmadi-L\'{o}pez $^{1,2,3,4}$}
\email{mariam.bouhmadi@ehu.eus}

\author{Che-Yu Chen $^{5,6}$ }
\email{b97202056@gmail.com}

\author{Pisin Chen $^{5,6,7}$}
\email{pisinchen@phys.ntu.edu.tw}

\date{\today}

\affiliation{
${}^1$Departamento de F\'{\i}sica, Universidade da Beira Interior, Rua Marqu\^{e}s D'\'Avila e Bolama 6200-001 Covilh\~a, Portugal\\
${}^2$Centro de Matem\'atica e Aplica\c{c}\~oes da Universidade da Beira Interior, Rua Marqu\^{e}s D'\'Avila e Bolama 6200-001 Covilh\~a, Portugal\\
${}^3$Department of Theoretical Physics University of the Basque Country UPV/EHU. P.O. Box 644, 48080 Bilbao, Spain\\
${}^4$IKERBASQUE, Basque Foundation for Science, 48011, Bilbao, Spain\\
${}^5$Department of Physics and Center for Theoretical Sciences, National Taiwan University, Taipei, Taiwan 10617\\
${}^6$LeCosPA, National Taiwan University, Taipei, Taiwan 10617\\
${}^7$Kavli Institute for Particle Astrophysics and Cosmology, SLAC National Accelerator Laboratory, Stanford University, Stanford, CA 94305, USA\\
}

\begin{abstract}
We study the quantum avoidance of the big rip singularity in the Eddington-inspired-Born-Infeld (EiBI) phantom model. Instead of considering a simple phantom dark energy component, which is described by a perfect fluid, we consider a more fundamental degree of freedom corresponding to a phantom scalar field with its corresponding potential, which would lead the classical universe to a big rip singularity. We apply a quantum geometrodynamical approach by performing an appropriate Hamiltonian study including an analysis of the constraints of the system. We then derive the Wheeler-DeWitt (WDW) equation and see whether the solutions to the WDW equation satisfy the DeWitt boundary condition. We find that by using a suitable Born-Oppenheimer (BO) approximation, whose validity is proven, the DeWitt condition is satisfied. Therefore, the big rip singularity is expected to be avoided in the quantum realm.
\end{abstract}



\maketitle
\section{Introduction}
It is well known that Einstein's general relativity (GR) predicts the existence of spacetime singularities in several physical configurations. In Refs.~\cite{Penrose:1964wq,LSSST,Hawking:1969sw}, it has been proven that spacetime singularities, which are accompanied with an abrupt termination of timelike (lightlike) geodesics, exist as long as the matter fields satisfy the strong (null) energy condition. This is the definition of spacetime singularity based on the notion of the geodesic incompleteness. In this regard, the \textit{attractive forces} between matter fields ensure the convergence of the geodesic congruence. At a certain spacetime point, the geodesics cannot be further extended and this is where the spacetime singularity is formally defined. For instance, GR predicts the existence of black hole singularities as well as the big bang singularity at the very beginning of the universe. 

Moreover, one should be aware that spacetime singularities are not restricted to small scales, they could in fact exist at large scales as well. In order to explain the late time accelerating expansion of the universe, one may consider a universe filled with some kinds of dark energy violating not only the strong energy condition, but also the null energy condition. This particular dark energy is called phantom dark energy and it could result in several cosmic curvature singularities in the future of the universe \cite{Nojiri:2005sx}. The most famous cosmic singularity associated with phantom dark energy is the big rip singularity \cite{Starobinsky:1999yw,Caldwell:1999ew,Carroll:2003st,Caldwell:2003vq,Chimento:2003qy,Dabrowski:2003jm,GonzalezDiaz:2003rf,GonzalezDiaz:2004vq,BouhmadiLopez:2006fu,Bamba:2012cp,BouhmadiLopez:2004me}. The size as well as the curvature of the universe diverge at the singularity and in a finite cosmic time. Before the universe reaches the singularity where the spacetime would be destroyed, all bounded structures would be ripped asunder by the strong Hubble flow. In addition to the big rip singularity, there exist several cosmic singularities and abrupt events that can be driven by phantom dark energy \cite{Barrow:2004xh,BouhmadiLopez:2007qb,Nojiri:2005sr,Bamba:2008ut,Frampton:2011sp,Brevik:2011mm,Bouhmadi-Lopez:2014cca,BouhmadiLopez:2005gk}.

In order to ameliorate the aforementioned spacetime singularities, one may resort to some quantum effects expected near the classical singularities. One of the promising approaches to address this issue is based on the quantum geometrodynamics in which the Wheeler-DeWitt (WDW) equation describes the quantum state of the universe as a whole \cite{qgkiefer}. The WDW equation in GR is derived from the Hamiltonian constraint defined by the Einstein equation. If the solution to the WDW equation satisfies the DeWitt (DW) boundary condition \cite{DeWitt:1967yk}, that is, it vanishes at the configuration corresponding to the classical singularity, one may claim that the singularity is then expected to be avoidable when quantum effects are taken into account. Following this direction, it has been shown in Refs.~\cite{Dabrowski:2006dd,Alonso-Serrano:2018zpi} that the big rip singularity in GR can be removed by quantum effects. The quantum behaviors of other cosmological singularities, such as the sudden singularity \cite{Kamenshchik:2007zj}, the big freeze singularity \cite{BouhmadiLopez:2009pu}, the type IV singularity \cite{Bouhmadi-Lopez:2013tua}, the little rip \cite{Albarran:2016ewi}, and the little sibling of the big rip \cite{Albarran:2015cda}, have also been discussed in the literature. It should be noticed that in some particular cases as shown in Refs.~\cite{Barvinsky:1993jf,Kamenshchik:2012ij,Barvinsky:2013aya}, a vanishing wave function at the classical singularity does not imply a vanishing probability because the measure used to define the probability diverges in those cases.

On the other hand, one can also address the spacetime singularities from a somewhat phenomenological point of view. Since a well-constructed and complete quantum theory of gravity is still lacking, one can consider some extended theories of gravity which not only reduce to GR in some proper limits, but also ameliorate spacetime singularities to some extent \cite{Capozziello:2011et}. These theories of gravity can also be regarded as effective theories of an unknown quantum theory of gravity under some cutoff energy scales. Within this scope an interesting theory is the Eddington-inspired-Born-Infeld (EiBI) gravity \cite{Banados:2010ix}. This theory contains higher order correction terms of curvature as compared with GR and it removes the big bang singularity \cite{Scargill:2012kg,Cho:2012vg,Avelino:2012ue}. In addition, the EiBI cosmology \cite{Delsate:2012ky,EscamillaRivera:2012vz,Yang:2013hsa,Du:2014jka,Cho:2013pea,Li:2017ttl}, the astrophysical configurations \cite{Olmo:2013gqa,Wei:2014dka,Sotani:2014lua,Sotani:2015ewa,Chen:2018mkf,Harko:2013wka,Sham:2013cya,Roshan:2018pts,Shaikh:2018yku}, constraints of the parameters in the theory \cite{Casanellas:2011kf,Avelino:2012ge,Avelino:2012qe,Jana:2017ost,Latorre:2017uve}, and some generalizations of the EiBI theory \cite{Liu:2012rc,Makarenko:2014lxa,Fernandes:2014bka,Odintsov:2014yaa,Jimenez:2014fla,Chen:2015eha,Bouhmadi-Lopez:2017lbx,Chen:2017ify,Ping-Li:2018mrt} have been widely investigated in the literature (see Ref.~\cite{BeltranJimenez:2017doy} for a nice review of the EiBI gravity).  

Even though the big bang singularity can be avoided in the EiBI theory, in our previous works \cite{Bouhmadi-Lopez:2013lha,Bouhmadi-Lopez:2014jfa} we have shown that the big rip singularity still exists in the classical theory. It is then natural to ask whether the big rip singularity in the EiBI theory can be alleviated by quantum effects. In Refs.~\cite{Bouhmadi-Lopez:2016dcf,Bouhmadi-Lopez:2018tel}, we considered a quantum geometrodynamical approach to address this issue by modeling the phantom dark energy as a perfect fluid. We have proven the existence of wave functions satisfying the DeWitt boundary condition near the configuration of the big rip and have therefore shown the possibility that the singularity can be avoided by quantum effects. In addition, some future abrupt events driven by phantom dark energy, such as the little rip and the little sibling of the big rip, have also been found avoidable under quantum effects in the EiBI model \cite{Albarran:2017swy}. For another quantum treatment for the EiBI gravity,  we refer to the scrutiny about the quantum tunneling effects of the singular instanton \cite{Arroja:2016ffm} and the regular instanton \cite{Bouhmadi-Lopez:2018sto}. One of the important results is the formation of a Lorentzian wormhole during bubble materialization shown in Ref.~\cite{Bouhmadi-Lopez:2018sto}.

In this paper, we consider a more general matter content than the phantom perfect fluid to study the quantum avoidance of the big rip singularity within the EiBI model. Specifically, we introduce a proper degree of freedom related to the matter sector described by a (phantom) scalar field $\phi$ minimally coupled to EiBI gravity to see whether a wave function satisfying the DeWitt condition is attainable. In a homogeneous and isotropic universe, there are two dynamical degrees of freedom in the system, one from the geometry side and the other from the matter side. We will start obtaining, for the first time, a consistent WDW equation for a universe filled with a minimally coupled scalar field within the EiBI setup. We will prove that under a suitable Born-Oppenheimer (BO) approximation, a wave function satisfying the DeWitt condition can be obtained. Therefore, the big rip singularity is expected to be avoided by quantum effects.   

This paper is outlined as follows. In section \ref{secII}, we concisely review the classical EiBI phantom model and exhibit how a big rip singularity would occur by adding a phantom scalar field with a proper potential. In section \ref{secIII}, we construct the modified WDW equation by considering an alternative action in Einstein frame. The model corresponds to a constrained system and a thorough Hamiltonian analysis as well as the quantization with the Dirac brackets are performed. In section \ref{secIV}, we solve the WDW equation by applying a Born-Oppenheimer approximation and show that the big rip singularity is hinted to be avoided according to the DeWitt criterion. We finally present our conclusions in section \ref{seccon}.

\section{The EiBI model with a phantom scalar field: the big rip singularity}\label{secII}
The EiBI action, which was proposed in Ref.~\cite{Banados:2010ix}, is 
\begin{equation}
S_{EiBI}=\frac{2}{\kappa}\int d^4x\Big[\sqrt{|g_{\mu\nu}+\kappa R_{(\mu\nu)}(\Gamma)|}-\lambda\sqrt{-g}\Big]+S_M(g),
\label{actioneibi}
\end{equation}
where $S_M$ is the matter Lagrangian coupled only to the physical metric $g_{\mu\nu}$. The theory is assumed to contain only the symmetric part of the Ricci tensor $R_{(\mu\nu)}$ and the curvature is constructed by the affine
connection $\Gamma$, which is independent of $g_{\mu\nu}$. Within this setup, the theory respects the projective gauge symmetry and the torsion field, if it exists, can be removed by simply choosing a gauge. On the other hand, the dimensionless constant $\lambda$ quantifies the effective cosmological constant at
the low curvature limit. Moreover, on the action \eqref{actioneibi} $|g_{\mu\nu}+\kappa R_{(\mu\nu)}|$ stands for the absolute value of the determinant of the rank two tensor $g_{\mu\nu}+\kappa R_{(\mu\nu)}$. Finally, $\kappa$ characterizes the theory and has inverse dimensions to that of a cosmological constant. In addition, we assume $8\pi G=c=1$.

The EiBI theory is equivalent to GR in vacuum, while it could deviate from GR when matter fields are included. In the early universe, the theory has been shown to be free of the big bang singularity. Furthermore, it should be stressed that the equations of motion of the theory contain derivatives of the metric up to only second order because of the Palatini structure of the theory. To be more precise, one can define an auxiliary metric $\lambda q_{\mu\nu}=g_{\mu\nu}+\kappa R_{(\mu\nu)}$ such that $q_{\mu\nu}$ is compatible with the connection. One of the two field equations relates algebraically the matter field with the two metrics, and the other equation corresponds to a second order differential equation of $q_{\mu\nu}$.

Even though the EiBI gravity is characterized by its ability to avoid cosmological singularity of the big bang types, it has been shown in Refs.~\cite{Bouhmadi-Lopez:2013lha,Bouhmadi-Lopez:2014jfa} that the big rip singularity is still unavoidable in the EiBI phantom model. Considering a homogeneous and isotropic spacetime which can be described by the following metric ansatz:
\begin{equation}
ds_g^2=-N(t)^2dt^2+a(t)^2\delta_{ij}dx^idx^j\,,\qquad
ds_q^2=-M(t)^2dt^2+b(t)^2\delta_{ij}dx^idx^j\,,
\end{equation}
where $N(t)$ and $a(t)$ are the lapse function and the scale factor of the physical metric $g_{\mu\nu}$, while $M(t)$ and $b(t)$ are the lapse function and the scale factor of the auxiliary metric $q_{\mu\nu}$. In this metric ansatz, these four quantities can be expressed as functions of the cosmic time $t$ and their evolutions in time are determined by the Euler-Lagrange equations of motion. 

In Refs.~\cite{Bouhmadi-Lopez:2013lha,Bouhmadi-Lopez:2014jfa}, we have demonstrated that in the EiBI gravity, if the universe is dominated by a phantom dark energy with a constant equation of state $w<-1$, the universe would end up at a big rip singularity. In fact, the asymptotic behavior of the Hubble rate (defined through the physical metric) near the singularity can be expressed as
\begin{equation}
\frac{1}{N^2}H^2\equiv\frac{1}{N^2}\Big(\frac{\dot{a}}{a}\Big)^2\approx\frac{4\sqrt{|w|^3}}{3(3w+1)^2}\rho_0a^{-3(1+w)},
\label{Friedmanneq}
\end{equation}
when $a$ goes to infinity, where $\rho_0$ is the energy density of the phantom dark energy at present time. The dot denotes the cosmic time derivative. Therefore, $H$ blows up which can be shown to be at finite cosmic time form now. Besides, the cosmic time derivative of $H$ diverges as well when approaching the big rip. 

If the phantom dark energy considered in this setup is governed by a phantom scalar field $\phi$ and its potential $V(\phi)$, the energy density and pressure of this phantom scalar field can be written as{\footnote{We will not consider dark matter on our approach because its effect is negligible in the asymptotic regime when the big rip singularity is reached. Furthermore, we will only consider the case where $\kappa$ is positive because of the instability problems usually present if $\kappa<0$ \cite{Avelino:2012ge,Avelino:2012qe}.}}
\begin{align}
\rho_\phi&=-\frac{\dot{\phi}^2}{2N^2}+V(\phi)\,,\label{rhophi}\\
p_\phi&=-\frac{\dot{\phi}^2}{2N^2}-V(\phi)\,.\label{pphi}
\end{align}
When the universe is approaching the big rip singularity, the energy density and pressure can be approximated as $\rho_\phi\approx\rho_0a^{-3(1+w)}$ and $p_\phi\approx w\rho_0a^{-3(1+w)}$. Combining Eqs.~\eqref{Friedmanneq}, \eqref{rhophi} and \eqref{pphi}, one can obtain the asymptotic form of the scalar field as a function of $a$:
\begin{equation}
\phi(a)=\pm\sqrt{\frac{3|w+1|(3w+1)^2}{4\sqrt{|w|^3}}}\ln{a},
\end{equation}
and the scalar field potential
\begin{equation}
V(\phi)=A\textit{e}^{B|\phi|},
\label{Vphi}
\end{equation}
where 
\begin{equation}
A=\frac{\rho_0}{2}(1-w)\,,
\qquad
B=\sqrt{\frac{12|w+1|\sqrt{|w|^3}}{(3w+1)^2}}\,.
\end{equation}
It can be seen that the big rip singularity is accompanied by a divergence of the scalar field, i.e., $|\phi|\rightarrow\infty$ and its potential.

\section{From the classical realm to the quantum world: derivation of the Wheeler-DeWitt equation}\label{secIII}
\subsection{The classical Hamiltonian}
To derive the WDW equation, it is necessary to obtain the correct classical Hamiltonian $\mathcal{H}_T$, which gives the classical equations of motion, and then to promote the Hamiltonian to a quantum operator: $\mathcal{H}_T\rightarrow\hat{\mathcal{H}}_T$. The WDW equation is essentially a restriction on the Hilbert space that operates over the wave function of the universe $|\Psi\rangle$, more precisely, $\hat{\mathcal{H}}_T|\Psi\rangle=0$. 

In Refs.~\cite{Delsate:2012ky,BeltranJimenez:2017doy}, it has been shown that the original EiBI action can be transformed to its Einstein frame through a proper Legendre transformation. The new action reads
\begin{equation}
\mathcal{S}_a=\lambda\int d^4x\sqrt{-q}\Big[R(q)-\frac{2\lambda}{\kappa}+\frac{1}{\kappa}\Big(q^{\alpha\beta}g_{\alpha\beta}-2\sqrt{\frac{g}{q}}\Big)\Big]+S_M(g),\label{action2}
\end{equation}
in which the fundamental variables become $g_{\mu\nu}$ and the auxiliary metric $q_{\mu\nu}$. The equations of motion obtained from the original action \eqref{actioneibi} can be reproduced by varying the above action \eqref{action2} with respect to $g_{\mu\nu}$ and $q_{\mu\nu}$. In Refs.~\cite{Bouhmadi-Lopez:2016dcf,Albarran:2017swy,Bouhmadi-Lopez:2018tel}, it has also been shown that the classical Hamiltonian is easier to define from this action because $\mathcal{S}_a$ resembles the standard Einstein-Hilbert action, with which the derivation of the Hamiltonian and the WDW equation is well-known. We then consider the action \eqref{action2} and assume that the matter sector is governed by a scalar field $\phi$ with its potential $V(\phi)$. After integrations by part, the reduced Lagrangian constructed from the action, $S_a=v_0\int dt\mathcal{L}$, can be rewritten as
\begin{equation}
\mathcal{L}=\lambda Mb^3\Big[-\frac{6{\dot{b}}^2}{M^2b^2}-\frac{2\lambda}{\kappa}+\frac{1}{\kappa}\Big(\frac{N^2}{M^2}+3\frac{a^2}{b^2}-2\frac{Na^3}{Mb^3}\Big)\Big]+Na^3\Big(l\frac{\dot{\phi}^2}{N^2}-2V(\phi)\Big),
\label{LA}
\end{equation}
where $v_0$ corresponds to the spatial volume after a proper compactification for spatially flat sections. On the above equation $l=\pm1$ denotes the ordinary scalar field ($+1$) and phantom scalar field ($-1$), respectively. For the sake of later convenience, we use the following changes of variables
\begin{equation}
X\equiv\frac{N}{M}\,,\qquad Y\equiv\frac{a}{b}\,,
\end{equation}
to replace $N$ and $a$ with $X$ and $Y$, respectively. The Lagrangian then becomes
\begin{equation}
\mathcal{L}=\lambda Mb^3\Big[-\frac{6{\dot{b}}^2}{M^2b^2}-\frac{2\lambda}{\kappa}+\frac{1}{\kappa}(X^2+3Y^2-2XY^3)\Big]+MXb^3Y^3\Big(l\frac{\dot{\phi}^2}{M^2X^2}-2V(\phi)\Big).
\label{LA2}
\end{equation}

The conjugate momenta of this system are
\begin{align}
&\,p_b=-\frac{12\lambda b}{M}\dot{b}\,,\qquad p_\phi=\frac{2b^3Y^3}{MX}l\dot{\phi}\,,\\
&\,p_X=0\,,\quad p_Y=0\,,\quad p_M=0\,.
\end{align}
It can be seen that the variables $\dot{X}$, $\dot{Y}$, and $\dot{M}$ cannot be inverted to be functions of canonical variables and their conjugate momenta, so the system is a constrained system and it has three primary constraints \cite{Diraclecture}:
\begin{equation}
p_X\sim0\,,\quad p_Y\sim0\,,\quad p_M\sim0\,,
\end{equation}
where $\sim$ denotes the weak equality, i.e., an equality on the constraint surface on which all the constraints in the phase space are satisfied. Provided the existence of the primary constraints, the Hamiltonian is defined as follows \cite{Diraclecture}
\begin{equation}
\mathcal{H}_T=-\frac{M}{24\lambda b}p_b^2+\frac{MX}{4b^3Y^3}lp_{\phi}^2-\frac{\lambda Mb^3}{\kappa}(X^2+3Y^2-2XY^3-2\lambda)+2MXb^3Y^3V(\phi)+\lambda_Mp_M+\lambda_Xp_X+\lambda_Yp_Y,
\label{HT}
\end{equation}
where $\lambda_M$, $\lambda_X$, and $\lambda_Y$ are Lagrange multipliers associated with each primary constraint. Note that the primary constraints are obtained directly from the definition of the conjugate momenta. These constraints should be satisfied throughout time and this would lead to the so called secondary constraints. To derive these secondary constraints requires the use of the equations of motion: $\dot{p}_M=[p_M,\mathcal{H}_T]\sim0$, $\dot{p}_X=[p_X,\mathcal{H}_T]\sim0$, and $\dot{p}_Y=[p_Y,\mathcal{H}_T]\sim0$. We call these requirements consistent conditions of the corresponding constraints and the consistent conditions of the primary constraints lead to the following secondary constraints
\begin{align}
C_X&\equiv\frac{\lambda X}{Y^3}-\lambda-\kappa\Big(\frac{l p_{\phi}^2}{8b^6Y^6}+V(\phi)\Big)\sim0\,,\label{CX}\\
C_Y&\equiv\frac{\lambda}{XY}-\lambda+\kappa\Big(\frac{l p_{\phi}^2}{8b^6Y^6}-V(\phi)\Big)\sim0\,,\label{CY}\\
C_M&\equiv\frac{p_b^2}{24\lambda b}-\frac{X}{4b^3Y^3}lp_{\phi}^2+\frac{\lambda b^3}{\kappa}(X^2+3Y^2-2XY^3-2\lambda)-2Xb^3Y^3V(\phi)\sim0\,.
\end{align}
Moreover, the consistent conditions of these secondary constraints do not generate new constraints anymore. Therefore, we have six constraints in this system. 

With these six constraints at hand, it is rather crucial to identify whether they are first class or second class \cite{Diraclecture}. By calculating the Poisson brackets between these constraints, it can be shown that $p_M$ is a first class constraint in the sense that its Poisson brackets with other constraints are zero weakly. On the other hand, $p_X$, $p_Y$, $C_X$, $C_Y$ and $C_M$ are second class constraints because they have at least one non-vanishing Poisson bracket with the other constraints on shell. Furthermore, the Hamiltonian $\mathcal{H}_T$ is also a constraint since it can be written as a linear combination of the constraints as follows
\begin{equation}
\mathcal{H}_T=-MC_M+\lambda_Mp_M+\lambda_Xp_X+\lambda_Yp_Y.
\label{hamit}
\end{equation}
Hence $\mathcal{H}_T\sim0$ is a first class constraint because its Poisson brackets with all the constraints should vanish weakly by definition. In the following subsection, we will use the Hamiltonian to write down the WDW equation, i.e., $\hat{\mathcal{H}}_T|\Psi\rangle=0$.

Essentially, the first class constraint $p_M$ corresponds to a gauge degree of freedom that can be fixed. An appropriate gauge condition is $M=\textit{constant}$ in the sense that $p_M$ and $M$ become second class constraints after choosing this gauge. The consistent condition of this gauge choice, that is, $[M,\mathcal{H}_T]\sim0$, implies $\lambda_M=0$.

\subsection{Quantization with Dirac brackets}
The system that we are dealing with contains several second class constraints. According to Ref.~\cite{Diraclecture} it was suggested that to quantize such a system, one needs to use the Dirac bracket, instead of the Poisson bracket, to define the commutation relations and promote the phase space functions to quantum operators. The Dirac bracket is basically constructed by calculating the Poisson brackets among all independent second class constraints of the system. The notion of \textit{independent} second class constraints means that one can not obtain any other first class constraints by taking linear combinations of these second class constraints. The independent second class constraints in our system (after choosing the gauge) are $\chi_i=\{M,\,p_M,\,p_X,\,p_Y,\,C_X,\,C_Y\}$. Note that the Hamiltonian \eqref{hamit} is a first class constraint and it can be written as a linear combination containing $C_M$. That is the reason why we have excluded $C_M$ when defining $\chi_i$.

The Dirac bracket of two phase space functions $F$ and $G$ is defined by \cite{Diraclecture} 
\begin{equation}
[F,G]_D=[F,G]-[F,\chi_i]\Delta_{ij}[\chi_j,G],
\end{equation}
where $\Delta_{ij}$ is a matrix satisfying
\begin{equation}
\Delta_{ij}[\chi_j,\chi_k]=\delta_{ik}.
\end{equation}
One of the important properties of the Dirac bracket is that the Dirac bracket of a second class constraint with any phase space function is zero strongly, i.e., $[\chi_i,G]_D=0$. This means that after promoting the phase space functions to quantum operators via the Dirac bracket, the second class constraints $\chi_i$ can be treated as zero operators and the Hamiltonian can be significantly simplified. By considering the constraints $C_X$ and $C_Y$, the Hamiltonian operator $\hat{\mathcal{H}}_T$ only contains $\hat{b}$, $\hat{p}_b$, $\hat{\phi}$, and $\hat{p}_\phi$. The Dirac brackets of the fundamental variables corresponding to these operators are
\begin{align}
[b,\phi]_D&=[b,p_\phi]_D=[\phi,p_b]_D=[p_b,p_\phi]_D=0,\nonumber\\
[b,p_b]_D&=[\phi,p_\phi]_D=1.\nonumber
\end{align}
This means that the standard commutation relations are still valid. In the next subsection we will make use of the second class constraints $\chi_i$ to derive an explicit form of the WDW equation.

\subsection{The WDW equation}
As mentioned previously, we can use the second class constraints $\chi_i$ to simplify the Hamiltonian then derive the WDW equation. If we combine the constraints \eqref{CX} and \eqref{CY}, we have
\begin{equation}
\lambda\Big(\frac{X}{Y^3}+\frac{1}{XY}\Big)=2\Big(\lambda+\kappa V(\phi)\Big).
\end{equation}
After inserting this equation into Eq.~\eqref{HT}, the Hamiltonian $\mathcal{H}_T$ can be simplified as follows
\begin{equation}
\mathcal{H}_T=M\Big(-\frac{p_b^2}{24\lambda b}+\frac{X}{Y^3}\frac{lp_\phi^2}{4b^3}+\frac{2\lambda b^3}{\kappa}(\lambda-Y^2)\Big)+\lambda_Xp_X+\lambda_Yp_Y.
\label{HT2}
\end{equation}
Note that $\lambda_M=0$ after the gauge fixing.

Next, we will use the asymptotic behaviors of the constraints \eqref{CX} and \eqref{CY} near the singularity to replace $X$ and $Y$ in Eq.~\eqref{HT2}. Near the big rip singularity, Eqs.~\eqref{CX} and \eqref{CY} can be approximated as ($l=-1$ for a phantom scalar field)
\begin{align}
\frac{X}{Y^3}&\approx\frac{\kappa}{\lambda}\rho_0a^{-3(1+w)}\approx c_1V(\phi),\\
Y^2&\approx\frac{\lambda a^{3(1+w)}}{\sqrt{|w|}\kappa\rho_0}\approx \frac{c_2}{V(\phi)},\label{3.18}
\end{align}
where $c_1$ and $c_2$ are two positive constants defined as
\begin{equation}
c_1\equiv\frac{2\kappa}{(1+|w|)\lambda}\,,\qquad c_2\equiv(\sqrt{|w|}c_1)^{-1}.
\end{equation}
Note that along the classical trajectory, we have $V(\phi(a))\approx Aa^{-3(1+w)}$ when approaching the singularity. The Hamiltonian can be written as
\begin{equation}
\mathcal{H}_T=M\Big[-\frac{p_b^2}{24\lambda b}-c_1V(\phi)\frac{p_\phi^2}{4b^3}+\frac{2\lambda b^3}{\kappa}\Big(\lambda-\frac{c_2}{V(\phi)}\Big)\Big]+\lambda_Xp_X+\lambda_Yp_Y,
\label{HT3}
\end{equation}
where the potential $V(\phi)$ is given by Eq.~\eqref{Vphi}.

We construct the WDW equation as follows
\begin{equation}
\langle b\phi|b^3\hat{\mathcal{H}}_T|\Psi\rangle=0,
\end{equation}
and choose, for simplicity, the following factor ordering
\begin{align}
b^2{\hat{p}_b}^2&=-\hbar^2\Big(b\frac{\partial}{\partial b}\Big)\Big(b\frac{\partial}{\partial b}\Big)=-\hbar^2\Big(\frac{\partial}{\partial x}\Big)\Big(\frac{\partial}{\partial x}\Big),\\
\textit{e}^{B|\phi|}{\hat{p}_\phi}^2&=-\hbar^2\Big(\textit{e}^{B|\phi|/2}\frac{\partial}{\partial\phi}\Big)\Big(\textit{e}^{B|\phi|/2}\frac{\partial}{\partial\phi}\Big)=-\hbar^2\Big(\frac{\partial}{\partial\tilde\phi}\Big)\Big(\frac{\partial}{\partial\tilde\phi}\Big).
\end{align}
Note that we have defined two new variables 
\begin{equation}
x=\ln{(\sqrt{\lambda}b)}\,,\qquad\tilde\phi=\mp\frac{2}{B}\textit{e}^{-\frac{B}{2}|\phi|},
\end{equation}
where the minus (plus) sign in $\tilde\phi$ corresponds to a positive (negative) $\phi$. Note as well that $\tilde\phi\rightarrow0$ when $|\phi|\rightarrow\infty$, i.e., close to the big rip singularity (on the classical trajectory, the relation between the two scale factors $a$ and $b$ is given in the footnote \ref{footnote2}). Finally, the WDW equation becomes
\begin{equation}
\Bigg[\frac{\hbar^2}{24}\partial_x^2+\frac{\hbar^2}{4}\kappa\rho_0\partial_{\tilde{\phi}}^2+\frac{2\textit{e}^{6x}}{\kappa}\Big(1-c_3{\tilde\phi}^2\Big)\Bigg]\Psi(x,\tilde\phi)=0,
\label{WDWfin}
\end{equation}
where
\begin{equation}
c_3\equiv\frac{f(w)}{\kappa \rho_0} \qquad \textrm{and} \qquad f(w)\equiv\frac{3|w(w+1)|}{(3w+1)^2}.
\end{equation}
The absence of $p_X$ and $p_Y$ is due to the fact that they can be treated as zero operators at the quantum level.

\section{The fulfillment of the DeWitt condition}\label{secIV}
 To solve the WDW equation \eqref{WDWfin}, we apply a BO approximation. This method gives an approximated solution where the total wave function can be written as 
\begin{equation}\label{ansatz}
\Psi(x,\tilde{\phi})=\sum_{k}C_k(x)\varphi_k(x,\tilde{\phi}).
\end{equation}
The BO approximation is valid as far as  the derivatives of $\varphi_k(x,\tilde{\phi})$ with respect to $x$ are negligible [see  appendix \ref{BOapp}]. Within this assumption, the WDW equation given in Eq.~\eqref{WDWfin} can be separated into the following two differential equations:
 \begin{align}
 &\left[\frac{\hbar^{2}}{24}\partial_{x}^2+\frac{2e^{6x}}{\kappa}-E_{k}\right]C_k(x)=0, \label{eqsep1}\\
 &\left[\frac{\hbar^2}{4}\kappa\rho_{0}\partial_{\tilde{\phi}}^2-\frac{2c_3}{\kappa }e^{6x}\tilde{\phi}^2+E_{k}\right]\varphi_k(x,\tilde{\phi})=0, \label{eqsep2} 
\end{align}
where $C_k(x)$  is the solution  to the gravitational part and $\varphi_k(x,\tilde{\phi})$ is the solution to the matter part. The decoupling constant  between  $C_k(x)$ and $\varphi_k(x,\tilde{\phi})$ is represented by the parameter $E_k$. We remind that the validity of the BO approximation should be justified once the solutions to the previous differential equations are obtained (we address this issue in the appendix \ref{BOapp}). 

Let us start with the gravitational part. The  solutions to the differential equation (\ref{eqsep1}) are given by (cf. Eq 9.1.54 of Ref. \cite{Abramow})
\begin{align}\label{gravpartsol}
C_k(x)=F_1 J_{\mu}\left[\alpha e^{3x}\right]+F_2 Y_{\mu}\left[\alpha e^{3x}\right],
\end{align}
where $F_1$ and $F_2$ are constants. On the other hand,  $J_{\mu}\left[\alpha e^{3x}\right]$ and  $Y_{\mu}\left[\alpha e^{3x}\right]$ are respectively, the first and the second kind Bessel functions of order $\mu$ and argument $\alpha e^{3x}$. The values of the parameters $\mu$ and $\alpha$ are 
\begin{align}
\mu&\equiv\frac{2\sqrt{6E_k}}{3\hbar},\label{mu}\\
\alpha&\equiv\frac{4\sqrt{3}}{3\hbar\kappa^{\frac{1}{2}}}.\label{alpha}
\end{align}
The asymptotic expressions for large arguments are given by (cf. Eq 9.2.1 and 9.2.2 of Ref. \cite{Abramow})
\begin{align}
 J_{\mu}\left[\alpha e^{3x}\right]\approx \sqrt{\frac{2}{\pi\alpha}}e^{-\frac{3}{2}x}\cos\left[\alpha e^{3x}-\frac{\pi}{4}(2\mu+1)\right], \label{Jasympt}\\
 Y_{\mu}\left[\alpha e^{3x}\right]\approx \sqrt{\frac{2}{\pi\alpha}}e^{-\frac{3}{2}x}\sin\left[\alpha e^{3x}-\frac{\pi}{4}(2\mu+1)\right]. \label{Yasympt}
\end{align}
As can be seen, both kinds of Bessel functions vanish at large scale factors. As the total wave function is a product of the gravitational and matter part solutions (cf Eq.~(\ref{ansatz})), it is possible to ensure the compliance of the DW boundary condition for an arbitrary $\tilde{\phi}$ as long as the matter part solution is bounded with respect to the metric variable $x$ when $x$ gets very large.

Let us solve the differential equation corresponding to the matter part. Notice that Eq. (\ref{eqsep2}) is the analogous to a Schr\"{o}dinger equation for a harmonic oscillator. However, in the present case, there are some remarkable differences: $i)$ Due to the phantom nature of the scalar field, the sign in the kinetic operator is switched, i.e., it becomes positive, while the potential is negative. However, the relative sign between them is the same. $ii)$ The field $ \tilde{\phi}$ can be positive or negative when reaching the big rip at  $|\tilde{\phi}|\rightarrow 0$. The solutions to the  Eq. (\ref{eqsep2}) can be  written as (cf. Ref.\cite{whittaker} chapters 16.5 and 16.51, pp .347-348)
\begin{equation}\label{matterpartsol}
\varphi_k(x,\tilde{\phi})=C_1 D_{\nu}\left[r(x)\tilde{\phi}\right]+C_2 D_{-\nu-1}\left[ir(x)\tilde{\phi}\right],
\end{equation}
where $C_1$ and $C_2$ are constants and $D_{\nu}[r(x)\tilde{\phi}]$ represents the parabolic cylinder function of order $\nu$ and argument $r(x)\tilde{\phi}$. The order and the argument of the parabolic cylinder function depend on the scale factor of the auxiliary metric through the variable $x$. Moreover, the parameter $\nu$ and the function $r(x)$ can be written as 
\begin{align}
&\nu=\pm\frac{\kappa^{\frac{1}{2}}E_k }{\sqrt{2f(w)}\hbar }e^{-3x}-\frac{1}{2}, \label{nuexp}\\
&r(x)=\beta e^{\frac{3}{2}x},\qquad \textrm{where}\qquad \beta\equiv\left[\frac{32f(w)}{\hbar^2\rho_0^2\kappa^3}\right]^{\frac{1}{4}}.\label{argexp}
\end{align}
The DW condition is applied close to the singularity, i.e. at very large values of the scale factor{\footnote{\label{footnote2}Recall that at the vicinity of the classical configuration corresponding to a big rip singularity, the auxiliary scale factor $b$ (or $x$) diverges because it can be related to the scale factor $a$ via Eq.~\eqref{3.18}
\begin{equation}
Y^2=\frac{a^2}{b^2}\approx\frac{c_2}{V(\phi)}\,,\qquad\Longrightarrow\qquad c_2b^2\approx a^2V(\phi)\rightarrow\infty\,,\nonumber
\end{equation}when $a\rightarrow\infty$ and $V(\phi)\rightarrow\infty$.} $a$}. Therefore, we focus on the limit $x\rightarrow\infty$. We will first consider the region where the argument $r(x)\tilde{\phi}$ is large. The reason of this priority is based on the classical trajectory. In fact, the classical trajectory is defined as
\begin{equation}\label{trajectory}
\tilde{\phi}(x)=\mp\frac{2}{B}\left[\sqrt{|w|}\kappa\rho_0\right]^{-\frac{h(w)}{2}}e^{h(w)x}\qquad \textrm{where} \qquad h(w)=-\frac{3(w+1)}{3w+1}.
\end{equation}
Since the value of the equation of state parameter is smaller than $-1$, the value of the exponent $h(w)$ is therefore negative. As expected, the classical limit of $\tilde{\phi}(x)$ tends to zero at large scale factors. Furthermore, the argument $r(x)\tilde{\phi}$ diverges classically since we have 
\begin{equation}\label{argument}
r(x)\tilde{\phi}\approx 
\exp\left\{\left[\frac{3}{2}+h(w)\right]x\right\}=\exp\left[\left(\frac{3w}{3w+1}\right)x\right],
\end{equation}
and $3w/(3w+1)$ is positive for $w<-1$. Hence, we will first focus on analyzing the asymptotic behavior of Eq.~(\ref{matterpartsol}) at large arguments, i.e., $r(x)\tilde{\phi}\gg1$. This also corresponds to a region in the parameter space where the metric variable $x$ is very large and $\tilde{\phi}$ is non-vanishing. Under this assumption, the order of the parabolic cylinder functions goes as $\nu\rightarrow-1/2$ while the argument  diverges, i.e. $r(x)\tilde{\phi}\rightarrow\infty$.
Therefore, the asymptotic behaviors of the solutions to the matter part read \cite{whittaker}
\begin{align}
D_{-\frac{1}{2}}\left[r(x)\tilde{\phi}\right]\approx& \  e^{-\frac{\left[r(x)\tilde{\phi}\right]^{2}}{4}}\left[r(x)\tilde{\phi}\right]^{-\frac{1}{2}}, \label{asymptoticD1}\\
D_{-\frac{1}{2}}\left[ir(x)\tilde{\phi}\right]\approx& \  e^{\frac{\left[r(x)\tilde{\phi}\right]^{2}}{4}}\left[r(x)\tilde{\phi}\right]^{-\frac{1}{2}}.\label{asymptoticD2}
\end{align} 
As the solution in Eq.~(\ref{asymptoticD2}) diverges very fast with respect to $x$, we select $C_2=0$ in order to ensure the compliance of the DW boundary condition. In addition, notice that for an order $\nu=-1/2$, the dependence on the decoupling constant $E_k$ disappears and the solution to the matter part simply reads
\begin{equation}\label{solgen}
\varphi_k(x,\tilde{\phi})= \frac{C_1}{\sqrt{\beta e^{\frac{3}{2}x}\tilde{\phi}}}\exp\left[-\frac{\beta^2e^{3x}\tilde{\phi}^{2}}{4}\right].
\end{equation}

Finally, the total wave function can be written as 
\begin{equation}\label{totalwave}
\Psi_{k}(x,\tilde{\phi})=C_k(x)\varphi_k(x,\tilde{\phi})=\left\{\tilde{C_1}J_{\mu}\left[\alpha e^{3x}\right]+\tilde{C_2}Y_{\mu}\left[\alpha e^{3x}\right]\right\}D_{\nu}\left[r(x)\tilde{\phi}\right],
\end{equation}
where $\tilde{C_1}$ and $\tilde{C_2}$ are constants and the asymptotic expression reads 
\begin{eqnarray}\label{totalwaveasympt}
\Psi_{k}(x,\tilde{\phi})\approx\sqrt{\frac{2}{\pi\alpha\beta\tilde{\phi}}}\exp\left[-\frac{9}{4}x-\frac{\beta^{2}}{4}(e^{3x}\tilde{\phi}^2)\right]\times \nonumber\\ 
\left\{\tilde{C_1}\cos\left[\alpha e^{3x}-\frac{\pi}{4}(2\mu+1)\right]+\tilde{C_2}\sin\left[\alpha e^{3x}-\frac{\pi}{4}(2\mu+1)\right]\right\}.
\end{eqnarray}
As we have found vanishing solutions at large scale factors for both the gravitational and the matter parts,  we can conclude that the total wave function vanishes at large scale factors and thus fulfills the DW boundary condition. 

On the other hand, we bear in mind that at the quantum level the classical relation $\tilde{\phi}(x)$ given in Eq.~(\ref{trajectory}) is meaningless and it just describes a curve where we expect to find confined the wave packet when approaching the semi-classical regime. There are regions in the parameter space, far away from the classical trajectory, where $\tilde{\phi}$ is small enough in such a way that the argument $r(x)\tilde{\phi}$ vanishes. For a vanishing argument $r(x)\tilde{\phi}$, the solution to the matter part given in  Eq.~(\ref{matterpartsol}) tends to a finite constant \cite{Abramow}. However, the total wave function is a product of the matter and the gravitational solutions, where the latter vanishes at large scale factors.
Therefore, once  the divergent solution on the matter part, Eq.~(\ref{asymptoticD2}), has been eliminated, it can be proven that for an arbitrary $\tilde{\phi}$ the total wave function vanishes at large scale factors.

\section{conclusion}\label{seccon}
In this paper, we have analyzed the quantum avoidance of the big rip singularity in the EiBI phantom model by using a quantum geometrodynamical approach. A complementary scrutiny has been carried out in our previous papers \cite{Bouhmadi-Lopez:2016dcf,Albarran:2017swy,Bouhmadi-Lopez:2018tel} in which the phantom dark energy was described by a perfect fluid for the sake of simplicity. In this paper, we consider a more general case in which the phantom dark energy is described by a phantom scalar field $\phi$ which adds an additional degree of freedom aside from the geometrical degree of freedom described by the scale factor of the universe.

In the quantum geometrodynamics, the WDW equation plays a very important role since it essentially describes the quantum behavior of the whole universe including the accompanied matter constituents. In order to obtain the WDW equation of this system, a correct and self-consistent Hamiltonian should be constructed at the classical level. One of the difficulties in the analysis of the model is from the affine structure of the EiBI theory since there are several additional auxiliary fields appearing in the system as compared with the GR case. However, in the end these additional auxiliary fields turn out to be reducible because of the corresponding second class constraints. We have identified all the first class and second class constraints in the system. More explicitly, one of the first class constraints $p_M$ actually corresponds to a gauge degree of freedom and it can be fixed by adding an additional constraint. The other first class constraint is, not surprisingly, the Hamiltonian and it has been used to construct the WDW equation. The existence of second class constraints requires a more careful treatment. In practice, one needs to use the Dirac bracket to promote the canonical variables of a system with second class constraints to quantum operators. It turns out that the WDW equation can be significantly simplified and it becomes a second order partial differential equation with two dynamical variables $x$ and $\tilde\phi$, as shown in Eq.~\eqref{WDWfin}.

To solve the WDW equation, obtained for the first time here for a minimally coupled scalar field in the EiBI theory of gravity, we impose a BO approximation on the wave function in the sense that the wave function can be decomposed into a gravitational part and a matter part. In the configuration where the classical big rip singularity is approached (large value of the scale factor $x$), we have proven that there exist solutions whose total wave function always satisfies the DeWitt boundary condition, no matter how the scalar field degree of freedom behaves. Therefore, the wave function vanishes near the configuration corresponding to the classical big rip singularity and the big rip singularity is thus hinted to be avoidable by quantum effects.  
\appendix 

\section{the Born Oppenheimer approximation}\label{BOapp}

The BO approximation considers that the back-reaction of the matter part on the gravitational part is negligible, i.e. the gravitational part varies faster with the scale factor than the matter part does, in such a way that 
 \begin{equation}\label{BOcond}
\frac{\partial^2\varphi_k(x,\tilde{\phi})}{\partial x^2}C_k(x),\frac{\partial\varphi_k(x,\tilde{\phi})}{\partial x}\frac{\partial C_k(x)}{\partial x}\ll \frac{\partial^2 C_k(x)}{\partial x^2}\varphi_k(x,\tilde{\phi}).
\end{equation}
This assumption leads to the separation of the initial equation (\ref{WDWfin}) into the coupled equations (\ref{eqsep1}) and (\ref{eqsep2}). Once these differential equations are solved, it is necessary to justify whether the obtained solutions $C(x)$ and $\varphi(x,\tilde{\phi})$ fulfill the inequality (\ref{BOcond}) previously imposed. In our results, the asymptotic behavior of the terms in Eq.~(\ref{BOcond}) can be written as 
\begin{align}
\frac{\partial^2 C_k}{\partial x^2}\varphi_k&\sim \frac{48}{\kappa\hbar^{2}}e^{6x}C_k\varphi_k, \label{terms1} \\
\frac{\partial\varphi_k}{\partial x}\frac{\partial C_k}{\partial x}&\sim \frac{24\sqrt{6}}{\hbar^{2}\kappa^{2}\rho_{0}}\left[f\left(w\right)\right]^{\frac{1}{2}}e^{6x}\tilde{\phi}^{2}C_k\varphi_k, \label{terms2}\\
\frac{\partial^2\varphi_k}{\partial x^2}C_k&\sim \frac{18f(w)}{\hbar^{2}\kappa^{3}\rho_{0}^2}e^{6x}\tilde{\phi}^{4}C_k\varphi_k\label{terms3}.
\end{align}
It should be stressed that although the parabolic cylinder functions obtained in Eq.~(\ref{matterpartsol}) have dependence with respect to the metric on both the order and on the argument, the order approaches a constant parameter when $x\rightarrow\infty$ (see Eq.~(\ref{nuexp})). Therefore, one can derive Eqs.~\eqref{terms1}, \eqref{terms2} and \eqref{terms3} by calculating the corresponding derivatives of the asymptotic expression obtained in Eq.~(\ref{asymptoticD1}). In this regard, the BO approximation method is valid as far as 
\begin{equation}\label{validcond}
\frac{\tilde{\phi}}{\sqrt{\kappa\rho_0}}\ll1.
\end{equation}
The term $\kappa\rho_0$ is dimensionless and it quantifies the current value of the phantom dark energy density normalized with the EiBI coupling constant $\kappa$. In any case, the inverse of $\tilde\phi$ should be very large (at least much larger than the EiBI scale and very close to the Planck scale) near the big rip singularity. This then justifies the BO approximation. In principle, the smaller the value of $\tilde{\phi}$ is, the better is the BO approximation. Despite that the BO method breaks down for large values of $\tilde{\phi}$, we can ensure that such an approach is still valid close to the region given by the classical trajectory, i.e., $x\rightarrow\infty$ and $\tilde{\phi}\rightarrow0$.

\acknowledgments
 
The work of IA was supported by a Santander-Totta fellowship ``Bolsas de Investiga\c{c}\~ao Faculdade de Ci\^encias (UBI) - Santander Totta". The work of MBL is supported by the Basque Foundation of Science Ikerbasque. She also would like to acknowledge the partial support from the Basque government Grant No. IT956-16 (Spain) and from the project FIS2017-85076-P (MINECO/AEI/FEDER, UE). CYC and PC are supported by Taiwan National Science Council under Project No. NSC 97-2112-M-002-026-MY3, Leung Center for Cosmology and Particle Astrophysics (LeCosPA) of National Taiwan University, and Taiwan National Center for Theoretical Sciences (NCTS). PC is in addition supported by US Department of Energy under Contract No. DE-AC03-76SF00515. CYC would like to thank the department of Theoretical Physics and History of Science of the University of the Basque Country (UPV/EHU) for kind hospitality while part of this work was done. Likewise, MBL is thankful to LeCosPA (National Taiwan University, Taipei) for kind hospitality while part of this work was done.


\begin{thebibliography}{99}

\bibitem{Penrose:1964wq}
  R.~Penrose,
  Phys.\ Rev.\ Lett.\  {\bf 14} (1965) 57.
  
\bibitem{Hawking:1969sw}
  S.~W.~Hawking and R.~Penrose,
  Proc.\ Roy.\ Soc.\ Lond.\ A {\bf 314} (1970) 529.
  
  \bibitem{LSSST}
S.~W.~Hawking, G.~F.~R.~Ellis,
{\em The Large Scale Structure of Space-Time}, Cambridge University Press, Cambridge, 1973.

\bibitem{Nojiri:2005sx}
  S.~Nojiri, S.~D.~Odintsov and S.~Tsujikawa,
  Phys.\ Rev.\ D {\bf 71} (2005) 063004.

\bibitem{Starobinsky:1999yw}
  A.~A.~Starobinsky,
  Grav.\ Cosmol.\  {\bf 6} (2000) 157.

\bibitem{Caldwell:1999ew}
  R.~R.~Caldwell,
  Phys.\ Lett.\ B {\bf 545} (2002) 23.

\bibitem{Carroll:2003st}
  S.~M.~Carroll, M.~Hoffman and M.~Trodden,
  Phys.\ Rev.\ D {\bf 68} (2003) 023509.

\bibitem{Caldwell:2003vq}
  R.~R.~Caldwell, M.~Kamionkowski and N.~N.~Weinberg,
  Phys.\ Rev.\ Lett.\  {\bf 91} (2003) 071301.

\bibitem{Chimento:2003qy}
  L.~P.~Chimento and R.~Lazkoz,
  Phys.\ Rev.\ Lett.\  {\bf 91} (2003) 211301.

\bibitem{Dabrowski:2003jm}
  M.~P.~D\c{a}browski, T.~Stachowiak and M.~Szyd{\l }owski,
  Phys.\ Rev.\ D {\bf 68} (2003) 103519.

\bibitem{GonzalezDiaz:2003rf}
  P.~F.~Gonz\'{a}lez-D\'{i}az,
  Phys.\ Lett.\ B {\bf 586} (2004) 1.

\bibitem{GonzalezDiaz:2004vq}
  P.~F.~Gonz\'{a}lez-D\'{i}az,
  Phys.\ Rev.\ D {\bf 69} (2004) 063522.

\bibitem{BouhmadiLopez:2004me}
  M.~Bouhmadi-L\'{o}pez and J.~A.~Jim\'{e}nez Madrid,
  JCAP {\bf 0505} (2005) 005.

\bibitem{BouhmadiLopez:2006fu}
  M.~Bouhmadi-L\'{o}pez, P.~F.~Gonz\'{a}lez-D\'{i}az and P.~Mart\'{i}n-Moruno,
  Phys.\ Lett.\ B {\bf 659} (2008) 1.

\bibitem{Bamba:2012cp}
  K.~Bamba, S.~Capozziello, S.~Nojiri and S.~D.~Odintsov,
  Astrophys.\ Space Sci.\  {\bf 342} (2012) 155.

\bibitem{Barrow:2004xh}
  J.~D.~Barrow,
  Class.\ Quant.\ Grav.\  {\bf 21} (2004) L79.

\bibitem{Nojiri:2005sr}
  S.~Nojiri and S.~D.~Odintsov,
  Phys.\ Rev.\ D {\bf 72} (2005) 023003.

\bibitem{BouhmadiLopez:2005gk}
  M.~Bouhmadi-L\'opez,
  Nucl.\ Phys.\ B {\bf 797} (2008) 78.

\bibitem{BouhmadiLopez:2007qb}
  M.~Bouhmadi-L\'opez, P.~F.~Gonz\'{a}lez-D\'{i}az and P.~Mart\'{i}n-Moruno,
  Int.\ J.\ Mod.\ Phys.\ D {\bf 17} (2008) 2269.

\bibitem{Bamba:2008ut}
  K.~Bamba, S.~Nojiri and S.~D.~Odintsov,
  JCAP {\bf 0810} (2008) 045.

\bibitem{Frampton:2011sp}
  P.~H.~Frampton, K.~J.~Ludwick and R.~J.~Scherrer,
  Phys.\ Rev.\ D {\bf 84} (2011) 063003.

\bibitem{Brevik:2011mm}
  I.~Brevik, E.~Elizalde, S.~Nojiri and S.~D.~Odintsov,
  Phys.\ Rev.\ D {\bf 84} (2011) 103508.

\bibitem{Bouhmadi-Lopez:2014cca}
  M.~Bouhmadi-L\'opez, A.~Errahmani, P.~Mart\'{i}n-Moruno, T.~Ouali and Y.~Tavakoli,
  Int.\ J.\ Mod.\ Phys.\ D {\bf 24} (2015) no.10,  1550078.

\bibitem{qgkiefer}
C.~Kiefer, \emph{Quantum Gravity}. Second edition (Oxford University Press, Oxford, 2007).

\bibitem{DeWitt:1967yk}
  B.~S.~DeWitt,
  Phys.\ Rev.\  {\bf 160} (1967) 1113.

\bibitem{Dabrowski:2006dd}
  M.~P.~D\c{a}browski, C.~Kiefer and B.~Sandh\"ofer,
  Phys.\ Rev.\ D {\bf 74} (2006) 044022.

\bibitem{Alonso-Serrano:2018zpi}
  A.~Alonso-Serrano, M.~Bouhmadi-L\'opez and P.~Mart\'in-Moruno, to appear in PRD,
  arXiv:1802.03290 [gr-qc].

\bibitem{Kamenshchik:2007zj}
  A.~Kamenshchik, C.~Kiefer and B.~Sandh\"ofer,
  Phys.\ Rev.\ D {\bf 76} (2007) 064032.

\bibitem{BouhmadiLopez:2009pu}
  M.~Bouhmadi-L\'opez, C.~Kiefer, B.~Sandh\"ofer and P.~Vargas Moniz,
  Phys.\ Rev.\ D {\bf 79} (2009) 124035.

\bibitem{Bouhmadi-Lopez:2013tua}
  M.~Bouhmadi-L\'opez, C.~Kiefer and M.~Kr\"amer,
  Phys.\ Rev.\ D {\bf 89} (2014) no.6,  064016.

\bibitem{Albarran:2016ewi}
  I.~Albarran, M.~Bouhmadi-L\'opez, C.~Kiefer, J.~Marto and P.~Vargas Moniz,
  Phys.\ Rev.\ D {\bf 94} (2016) no.6,  063536.

\bibitem{Albarran:2015cda}
  I.~Albarran, M.~Bouhmadi-L\'opez, F.~Cabral and P.~Mart\'in-Moruno,
  JCAP {\bf 1511} (2015) no.11,  044.

\bibitem{Barvinsky:1993jf}
  A.~O.~Barvinsky,
  Phys.\ Rept.\  {\bf 230} (1993) 237.

\bibitem{Kamenshchik:2012ij}
  A.~Y.~Kamenshchik and S.~Manti,
  Phys.\ Rev.\ D {\bf 85} (2012) 123518.

\bibitem{Barvinsky:2013aya}
  A.~O.~Barvinsky and A.~Y.~Kamenshchik,
  Phys.\ Rev.\ D {\bf 89} (2014) no.4,  043526.

\bibitem{Capozziello:2011et}
  S.~Capozziello and M.~De Laurentis,
  Phys.\ Rept.\  {\bf 509} (2011) 167.

\bibitem{Banados:2010ix}
  M.~B\~{a}nados and P.~G.~Ferreira,
  Phys.\ Rev.\ Lett.\  {\bf 105} (2010) 011101
   Erratum: [Phys.\ Rev.\ Lett.\  {\bf 113} (2014) no.11,  119901].

\bibitem{Cho:2012vg}
  I.~Cho, H.~C.~Kim and T.~Moon,
  Phys.\ Rev.\ D {\bf 86} (2012) 084018.

\bibitem{Scargill:2012kg}
  J.~H.~C.~Scargill, M.~Ba\~{n}ados and P.~G.~Ferreira,
  Phys.\ Rev.\ D {\bf 86} (2012) 103533.

\bibitem{Avelino:2012ue}
  P.~P.~Avelino and R.~Z.~Ferreira,
  Phys.\ Rev.\ D {\bf 86} (2012) 041501.

\bibitem{Delsate:2012ky}
  T.~Delsate and J.~Steinhoff,
  Phys.\ Rev.\ Lett.\  {\bf 109} (2012) 021101.

\bibitem{EscamillaRivera:2012vz}
  C.~Escamilla-Rivera, M.~Ba\~nados and P.~G.~Ferreira,
  Phys.\ Rev.\ D {\bf 85} (2012) 087302.

\bibitem{Cho:2013pea}
  I.~Cho, H.~C.~Kim and T.~Moon,
  Phys.\ Rev.\ Lett.\  {\bf 111} (2013) 071301.

\bibitem{Yang:2013hsa}
  K.~Yang, X.~L.~Du and Y.~X.~Liu,
  Phys.\ Rev.\ D {\bf 88} (2013) 124037.

\bibitem{Du:2014jka}
  X.~L.~Du, K.~Yang, X.~H.~Meng and Y.~X.~Liu,
  Phys.\ Rev.\ D {\bf 90} (2014) 044054.

\bibitem{Li:2017ttl}
  S.~L.~Li and H.~Wei,
  Phys.\ Rev.\ D {\bf 96} (2017) no.2,  023531.

\bibitem{Harko:2013wka}
  T.~Harko, F.~S.~N.~Lobo, M.~K.~Mak and S.~V.~Sushkov,
  Phys.\ Rev.\ D {\bf 88} (2013) 044032.

\bibitem{Sham:2013cya}
  Y.~H.~Sham, L.~M.~Lin and P.~T.~Leung,
  Astrophys.\ J.\  {\bf 781} (2014) 66.

\bibitem{Olmo:2013gqa}
  G.~J.~Olmo, D.~Rubiera-Garcia and H.~Sanchis-Alepuz,
  Eur.\ Phys.\ J.\ C {\bf 74} (2014) 2804.

\bibitem{Sotani:2014lua}
  H.~Sotani and U.~Miyamoto,
  Phys.\ Rev.\ D {\bf 90} (2014) 124087.

\bibitem{Wei:2014dka}
  S.~W.~Wei, K.~Yang and Y.~X.~Liu,
  Eur.\ Phys.\ J.\ C {\bf 75} (2015) 253
   Erratum: [Eur.\ Phys.\ J.\ C {\bf 75} (2015) 331].

\bibitem{Sotani:2015ewa}
  H.~Sotani and U.~Miyamoto,
  Phys.\ Rev.\ D {\bf 92} (2015) no.4,  044052.

\bibitem{Roshan:2018pts}
  M.~Roshan, A.~Kazemi and I.~De Martino,
  Monthly Notices of the Royal Astronomical Society, Volume 479,
  Issue 1, p.1287-1296, 2018.

\bibitem{Chen:2018mkf}
  C.~Y.~Chen and P.~Chen,
  Phys.\ Rev.\ D {\bf 98} (2018) no.4,  044042.

\bibitem{Shaikh:2018yku}
  R.~Shaikh,
  Phys.\ Rev.\ D {\bf 98} (2018) no.6,  064033.

\bibitem{Casanellas:2011kf}
  J.~Casanellas, P.~Pani, I.~Lopes and V.~Cardoso,
  Astrophys.\ J.\  {\bf 745} (2012) 15.

\bibitem{Avelino:2012ge}
  P.~P.~Avelino,
  Phys.\ Rev.\ D {\bf 85} (2012) 104053.

\bibitem{Avelino:2012qe}
  P.~P.~Avelino,
  JCAP {\bf 1211} (2012) 022.

\bibitem{Latorre:2017uve}
  A.~Delhom-Latorre, G.~J.~Olmo and M.~Ronco,
  Phys.\ Lett.\ B {\bf 780} (2018) 294.

\bibitem{Jana:2017ost}
  S.~Jana, G.~K.~Chakravarty and S.~Mohanty,
  Phys.\ Rev.\ D {\bf 97} (2018) no.8,  084011.

\bibitem{Liu:2012rc}
  Y.~X.~Liu, K.~Yang, H.~Guo and Y.~Zhong,
  Phys.\ Rev.\ D {\bf 85} (2012) 124053.

\bibitem{Makarenko:2014lxa}
  A.~N.~Makarenko, S.~Odintsov and G.~J.~Olmo,
  Phys.\ Rev.\ D {\bf 90} (2014) 024066.
  
\bibitem{Fernandes:2014bka}
  K.~Fernandes and A.~Lahiri,
  Phys.\ Rev.\ D {\bf 91} (2015) no.4,  044014.
  
\bibitem{Odintsov:2014yaa}
  S.~D.~Odintsov, G.~J.~Olmo and D.~Rubiera-Garcia,
  Phys.\ Rev.\ D {\bf 90} (2014) 044003.

\bibitem{Jimenez:2014fla}
  J.~Beltr\'an Jim\'enez, L.~Heisenberg and G.~J.~Olmo,
  JCAP {\bf 1411} (2014) 004.
  
\bibitem{Chen:2015eha}
  C.~Y.~Chen, M.~Bouhmadi-L\'opez and P.~Chen,
  Eur.\ Phys.\ J.\ C {\bf 76} (2016) 40.

\bibitem{Bouhmadi-Lopez:2017lbx}
  M.~Bouhmadi-L\'opez, C.~Y.~Chen and P.~Chen,
  JCAP {\bf 1711} (2017) no.11,  053.

\bibitem{Chen:2017ify}
  C.~Y.~Chen, M.~Bouhmadi-L\'opez and P.~Chen,
  Eur.\ Phys.\ J.\ C {\bf 78} (2018) no.1,  59.

\bibitem{Ping-Li:2018mrt}
  Ping-Li, J.~C.~Ding, Q.~Q.~Fan, X.~R.~Hu and J.~B.~Deng,
  arXiv:1808.00801 [gr-qc].

\bibitem{BeltranJimenez:2017doy}
  J.~Beltr\'an Jim\'enez, L.~Heisenberg, G.~J.~Olmo and D.~Rubiera-Garcia,
  Phys.\ Rept.\  {\bf 727} (2018) 1.

\bibitem{Bouhmadi-Lopez:2013lha}
  M.~Bouhmadi-L\'opez, C.~Y.~Chen and P.~Chen,
  Eur.\ Phys.\ J.\ C {\bf 74} (2014) 2802

\bibitem{Bouhmadi-Lopez:2014jfa}
  M.~Bouhmadi-L\'opez, C.~Y.~Chen and P.~Chen,
  Eur.\ Phys.\ J.\ C {\bf 75} (2015) 90.
  
\bibitem{Bouhmadi-Lopez:2016dcf}
  M.~Bouhmadi-L\'opez and C.~Y.~Chen,
  JCAP {\bf 1611} (2016) no.11,  023.
  
\bibitem{Bouhmadi-Lopez:2018tel}
  M.~Bouhmadi-L\'opez, C.~Y.~Chen and P.~Chen,
  arXiv:1810.10918 [gr-qc].

\bibitem{Albarran:2017swy}
  I.~Albarran, M.~Bouhmadi-L\'opez, C.~Y.~Chen and P.~Chen,
  Phys.\ Lett.\ B {\bf 772} (2017) 814.
  
\bibitem{Arroja:2016ffm}
  F.~Arroja, C.~Y.~Chen, P.~Chen and D.~h.~Yeom,
  JCAP {\bf 1703} (2017) no.03,  044.

\bibitem{Bouhmadi-Lopez:2018sto}
  M.~Bouhmadi-L\'opez, C.~Y.~Chen, P.~Chen and D.~h.~Yeom,
  JCAP {\bf 1810} (2018) no.10,  056.

\bibitem{Diraclecture}
  P.~A.~M.~Dirac, \textit{Lectures on Quantum Mechanics}, Yeshiva University, New York (1964).

  \bibitem{Abramow}
  M.~Abramowitz and I.~Stegun, \textit{Handbook on Mathematical Functions} (Dover, 1980).

    \bibitem{whittaker}
 E. T. Whittaker and G. N. Watson, \textit{A course of Modern Analysis}. Third edition, Cambridge University press (London, 1920).

  



 
 \end{thebibliography}
\end{document}